\documentclass[preprint,aps,pra,showpacs,floatfix]{revtex4-1}
\usepackage[utf8]{inputenc}
\usepackage[OT1]{fontenc}
\usepackage{amsmath}
\usepackage{amsfonts}
\usepackage{amssymb}
\usepackage{color}
\usepackage{graphicx}
\usepackage[left=3cm,right=1.5cm,top=2cm,bottom=2cm]{geometry}

\newcommand{\bfr}{{\bf r}}

\newcommand{\bfp}{{\bf p}}
\newcommand{\bfA}{{\bf A}}

\newcommand{\balpha}{{\mbox{\boldmath$\alpha$}}}

\newcommand{\be}{\begin{eqnarray}}
\newcommand{\ee}{\end{eqnarray}}
\newcommand{\la}{\langle}
\newcommand{\ra}{\rangle}

\newcommand{\veps}{\varepsilon}

\begin{document}

\title{Nuclear recoil effect on $g$ factor of heavy ions: prospects for tests of quantum electrodynamics in a new region}

\author{A. V. Malyshev, V. M. Shabaev, D. A. Glazov, I. I. Tupitsyn}

\affiliation{Department of Physics, St. Petersburg State University, Universitetskaya 7/9, 199034 St. Petersburg, Russia}

\begin{abstract}
The nuclear recoil effect on the $g$ factor of H- and Li-like heavy ions is evaluated
to all orders in $\alpha Z$. The calculations include an approximate treatment
of the nuclear size and the electron-electron interaction corrections to the recoil effect.
As the result, the second largest contribution to the theoretical uncertainty
of the $g$-factor values of $^{208}$Pb$^{79+}$ and $^{238}$U$^{89+}$
is strongly reduced.
Special attention is paid to tests of the QED recoil effect on the  $g$ factor 
in experiments  with heavy ions. 
It is found that, while the QED recoil effect on the  $g$-factor value
is masked by the uncertainties of the nuclear size and nuclear polarization
contributions, it can be probed on a few-percent level in 
the specific difference of the $g$ factors of H- and Li-like heavy ions.
This provides a unique opportunity to test QED in a new region 
--- strong-coupling regime beyond the Furry picture.  
\end{abstract}

\maketitle

\section{Introduction}

High-precision measurements of the Lamb shifts in highly charged heavy ions 
\cite{sto00,gum05,sch91,bra03,bei05} have already provided tests of QED 
effects at strong Coulomb field  on a few tenth percent level (see, e.g.,
Refs. \cite{gla11,vol13} and references therein).
To date, there are also a number of high-precision measurements of the hyperfine splitting (HFS)
in heavy H-like ions  \cite{kla94,cre96,cre98,see98,bei01,ull15}. 
The main goal of these experiments was to test QED in a combination of the strongest
electric and magnetic fields. However, due to a large uncertainty of the nuclear 
magnetization distribution correction (so-called Bohr-Weisskopf effect),  direct tests 
of QED by comparison of theory and experiment on the HFS in H-like ions
turned out to be impossible. The solution of the problem was found in Ref. \cite{sha01},
where it was proposed to study a specific difference of the HFS in H- and Li-like ions
of the same heavy isotope. This difference can be calculated to a very high accuracy. 
It took about 15 years to reach the required level of accuracy for the HFS measurements
in Li-like bismuth \cite{ull17}. 
A large discrepancy between the obtained experimental result  
and the most elaborated theoretical prediction \cite{volotka:12:prl} for the specific difference  has
established the ``hyperfine puzzle'' \cite{kar17}, which is presently 
a subject of intensive  investigations from both theoretical and experimental sides.  

The current tests of bound-state QED by studying the Lamb shift and the HFS in heavy ions are   
limited to the region where the standard formalism of quantum electrodynamics in presence of
 external classical fields can be applied. This formalism is known as the Furry
picture of QED. The nuclear recoil correction to the Lamb shift in heavy ions,
whose evaluation requires QED beyond the Furry picture, is
generally  masked by the uncertainties of the
nuclear size and polarization contributions. This fact and the limited
experimental accuracy 
prevent presently any precise tests of the QED recoil effect on the Lamb shifts
in heavy ions. 
Much higher accuracy is expected to be achieved in 
$g$-factor experiments with heavy ions which are anticipated in the 
nearest future at the HITRAP/FAIR facilities in Darmstadt and at the 
Max-Planck-Institut f\"ur Kernphysik (MPIK) 
in Heidelberg. To date, a number of high-precision measurements of the
$g$ factor was performed for low- and middle-$Z$ highly charged ions ($Z$ is the nuclear charge number) 
\cite{haf00,ver04,stu11,wag13,lin13,stu13,stu14,koel16}.
The measurement of the isotope shift of the $g$ factor of Li-like
$^{A}$Ca$^{17+}$ with $A=40$ and $A=48$   \cite{koel16}
has provided already the first test of the relativistic 
theory of the recoil effect in highly charged ions  
in the presence of magnetic field \cite{shaxx}. 
The precision of the experimental value  is currently limited by
the uncertainty of the  $A=48$ calcium atomic mass. 
Several worldwide initiatives
are presently aiming to improve 
the atomic masses. The accuracy improvement of the calcium masses will
result in a direct test of the QED recoil effect
in highly charged ions.
 Moreover, the $g$-factor experiments
for heavy ions, which are anticipated in the nearest future,
 should provide a unique possibility to test 
the QED recoil effect in strongly nonperturbative in 
$\alpha Z$ regime, provided the total theoretical
value is evaluated to the required accuracy. It is known \cite{sha02b,nef02}, 
however, that the uncertainty due to the nuclear size and polarization
effects grows strongly with $Z$ and masks the recoil 
effect for heavy ions. In Refs. \cite{sha:2002:062104,sha06,vol14,yerokhin:2016:100801} it was shown
that this uncertainty can be strongly reduced in specific differences
of the $g$-factor values of H-, Li-, and B-like ions allowing for more
precise tests of bound-state QED at strong fields.  In the present paper we 
evaluate the nuclear recoil effect on the $g$ factor of H- and Li-like $^{208}$Pb and 
$^{238}$U ions and demonstrate
that the QED recoil contribution to the specific difference in the 
case of Pb is two orders of magnitude bigger
than the uncertainty to which the total theoretical value of the difference can
be calculated. This will give a unique possibility to test QED at strong-coupling
regime beyond the Furry picture.  

The relativistic units ($\hbar=c=1$, $e<0$) are used in the paper.

\section{ Basic formulas}

The complete $\alpha Z$-dependent formula ($\alpha$ is the fine structure constant) for the nuclear recoil effect on the 
$g$ factor of a H-like ion to the first order in the electron-to-nucleus mass ratio
$m/M$ was derived in Ref. \cite{sha01a}. As was noted in that paper, the obtained formula
can partially account for the nuclear size correction to the recoil effect if
 the pure Coulomb potential  of the nucleus $V=-\alpha Z/r$ is replaced
with the potential of the extended nucleus. The replacement of the potential with
an effective local potential $V_{\rm eff}(r)$, which is the sum of the nuclear 
and screening potentials, allows one to partially account for the 
corrections to the one-electron recoil contribution due to the screening of the valence $2s$ electron 
by the closed $1s^2$ shell in a case of Li-like ion.
The $m/M$ one-electron nuclear recoil contribution to the $g$ factor for the  state $a$
can be represented by the sum of the low-order and  higher-order in $\alpha Z$ terms,
$\Delta g=\Delta g_{\rm L}+\Delta g_{\rm H}$, where 
\be \label{low}
\Delta g_{\rm L}&=&\frac{1}{\mu_0 {\cal H}  m_a}
\frac{1}{M} \la \delta a|
\Bigr[ \bfp^2
 -\frac{\alpha Z}{r}\Bigl(\balpha+\frac{(\balpha\cdot\bfr)\bfr}{r^2}\Bigr)
  \cdot\bfp
  \Bigr]|a\ra\nonumber\\
&& -\frac{1}{ m_a}\frac{m}{M}\la
a|\Bigl([\bfr \times \bfp]_z -\frac{\alpha Z}{2r}[\bfr \times \balpha ]_z\Bigr)|a\ra\,,
\ee
\be  \label{high}
\Delta g_{\rm H}&=&\frac{1}{\mu_0  {\cal H} m_a}
\frac{1}{M} \frac{i}{2\pi} \int_{-\infty}^{\infty} d\omega\;\bigg\{ \la
\delta a|\Bigl(D^k(\omega)-\frac{[p^k,V]}{\omega+i0}\Bigr)\nonumber\\
&&\times G(\omega+\veps_a)\Bigl(D^k(\omega)+\frac{[p^k,V]}{\omega+i0}\Bigr)|a\ra \nonumber\\
&&+\la a|\Bigl(D^k(\omega)-\frac{[p^k,V]}{\omega+i0}\Bigr)
G(\omega+\veps_a)\nonumber\\
&&\times\Bigl(D^k(\omega)+\frac{[p^k,V]}{\omega+i0}\Bigr) |\delta
a\ra\nonumber\\ &&+ \la a|\Bigl(D^k(\omega)-\frac{[p^k,V]}{\omega+i0}\Bigr)
G(\omega+\veps_a)(\delta V-\delta \veps_a)\nonumber\\ &&\times G(\omega+\veps_a)
\Bigl(D^k(\omega)+\frac{[p^k,V]}{\omega+i0}\Bigr)|a\ra \bigg\}\,. \label{eq3}
\ee
Here  $\mu_0$ is the Bohr magneton, $m_a$ is the angular momentum
projection of the state under consideration,
$p^k=-i\nabla^k$ is the momentum operator, 
$V(r)$ is the nuclear or effective potential (see the discussion above),
$|a\ra$ --- the Dirac wave function for the potential $V(r)$,
$\veps_a$ --- the corresponding Dirac energy, 
$\delta V(\bfr)=-e\balpha \cdot\bfA_{\rm cl}(\bfr)$
describes the interaction of the electron with the classical
homogeneous magnetic field  ${\bf A}_{\rm cl}(\bfr)=[{\bf {\cal H}}\times {\bfr}]/2$,
$G(\omega)=\sum_n|n\ra \la n|[\omega-\veps_n(1-i0)]^{-1}$, 
$\delta \veps_a=\la a|\delta V|a\ra$,  $|\delta a\ra=\sum_n^{\veps_n\ne
\veps_a}|n\ra\la n|\delta V|a\ra (\veps_a-\veps_n)^{-1}$,
$D^k(\omega)=-4\pi\alpha Z\alpha^l D^{lk}(\omega)$,
\begin{eqnarray} \label{06photon}
D^{lk}(\omega,{\bf r})&=&-\frac{1}{4\pi}\Bigl\{\frac
{\exp{(i|\omega|r)}}{r}\delta_{lk} +\nabla^{l}\nabla^{k}
\frac{(\exp{(i|\omega|r)}
-1)}{\omega^{2}r}\Bigr\}\,
\end{eqnarray}
is the transverse part of the photon propagator in the Coulomb 
gauge,  $\balpha$ is a vector of the Dirac matrices, and
the summation over the repeated indices is implied. 
The low-order contribution $\Delta g_{\rm L}$, which can be derived from the Breit equation,
we will refer to as the one-electron non-QED contribution.  
The higher-order  term  $\Delta g_{\rm H}$ is determined by quantum electrodynamics
beyond the Breit approximation and will be termed as the one-electron QED contribution. 

In the case of a few-electron ion, one should also consider the 
two-electron recoil contributions. For ions
with one electron over closed shells the two-electron contributions
can be easily taken into account within the framework of the one-electron approach 
by redefining the electron propagator \cite{sha02a,sha:2002:062104,shc15,shaxx}.  
For the ground state of a Li-like ion
the two-electron recoil term vanishes in the zeroth order in $1/Z$. 
The corresponding first- and higher-order corrections in $1/Z$ to the $g$ factor can be evaluated
employing the effective recoil operators derived
within the framework of the Breit approximation \cite{shaxx}.

\section{Results and discussion}

According to the aforesaid, the nuclear recoil correction to the $g$ factor is 
represented as a sum of the term corresponding to the Breit approximation and the QED term.
For the pure Coulomb field $V(r)=-\alpha Z/r$, 
the low-order (non-QED) one-electron term can be calculated analytically \cite{sha01}:
\begin{eqnarray} \label{06shabaeveq112}
\Delta g_{\rm L}({\rm p.n.})=-\frac{m}{M}
\frac{2\kappa^2\veps_a^2+\kappa m \veps_a-m^2}{2m^2j(j+1)}\,,
\end{eqnarray}
where $\veps_a$ is the Dirac energy and $\kappa=(-1)^{j+l+1/2}(j+1/2)$ is the
relativistic angular  quantum number.
In the present work the low-order (non-QED)  one-electron term is calculated
for extended nuclei using Eq.~(\ref{low}). 
The sums over the intermediate electron states have been evaluated using
the dual-kinetic-balance (DKB) finite basis set  method \cite{sha04}
with the basis functions constructed from B splines \cite{sap96}.
The results of the calculations for $Z=82, 92$
expressed in terms of the function $F(\alpha Z)$,
\be
\label{F}
\Delta g=\frac{m}{M}{(\alpha Z)^2} F(\alpha Z)\,,
\ee
are presented in Tables~\ref{H-like} and \ref{Li-like} for H- and Li-like ions, respectively. 
For H-like ions the results are presented for both  point and extended nuclei, 
while for Li-like ions only the case of extended nuclei  is considered.
The root-mean-square (rms) nuclear charge radii were taken from Ref. \cite{ang13}.
As mentioned above, for the ground state of a Li-like ion the two-electron
recoil contribution to the $g$ factor is equal to zero, if one considers the
independent electron approximation. This approximation corresponds to zeroth order in $1/Z$. 
In Table~\ref{Li-like} we present the two-electron recoil correction, which was evaluated
to all orders in $1/Z$  within the Breit approximation, as described in our recent work \cite{shaxx}.

In the point-nucleus case, the numerical calculations of the higher-order (QED) 
one-electron contribution $\Delta g_{\rm H}$
have been performed for the $1s$ and $2s$ states in Refs. \cite{sha02b,shaxx}.
In the present paper we have calculated this contribution for the extended nucleus case.
To partially account for the electron-electron interaction effect on the QED recoil contribution
for Li-like ions,  the core-Hartree (CH), Perdew-Zunger (PZ),
and local Dirac-Fock (LDF) potentials have been also employed.
The construction methods and application examples for these potentials
can be found in Refs.~\cite{per81,sap02,sha05,sap11,mal17}.  
The $\omega$ integration in Eq. (\ref{high})
was performed analytically for the  term which does not contain the $D^k(\omega)$ operator (``Coulomb'' term) 
and numerically, after the standard Wick's rotation, for the other 
(``one-transverse-photon'' and ``two-transverse-photon'') terms \cite{sha02b,shaxx}.  
The summation over the intermediate electron states
was carried out using the DKB finite basis set method.   
The results of the calculations for $Z=82, 92$
expressed  in terms of the function $F(\alpha Z)$, defined by Eq. (\ref{F}),  
are given in Tables~\ref{H-like} and \ref{Li-like}  for H- and Li-like ions, respectively.
It should be noted that in case of uranium ions the QED term is even bigger than the
non-QED contribution. 

For H-like ions (Table~\ref{H-like}), the uncertainty is mainly due to the approximate treatment
of the nuclear size contribution to the recoil effect on the $g$ factor.
We  assume that this uncertainty should be on the level of the related correction
to the binding energy which was studied in the Breit approximation in Ref. \cite{ale15}.
According to Ref. \cite{ale15}, this correction changes the  nuclear size contribution
to the recoil effect by 16\% and 21\% for the $1s$ state
of H-like lead and uranium ions, respectively. 

For Li-like ions (Table~\ref{Li-like}), 
as the final theoretical value of the QED recoil contribution, we have chosen the value obtained for the
LDF potential.  The  uncertainty is estimated as a sum of two contributions.
The first one is caused by the approximate treatment of the electron-electron
interaction corrections to the QED recoil effect. To estimate this uncertainty,
we have performed the calculations of the non-QED one-electron recoil contribution
with the LDF potential and compared the obtained results with the total non-QED recoil values 
presented in Table~\ref{Li-like}. 
The ratio of the difference obtained to the non-QED LDF result is chosen 
as a relative uncertainty of the corresponding correction to the
QED recoil contribution.
The second contribution to the uncertainty is 
due to the approximate treatment of the nuclear size contribution to the recoil effect.
It is estimated in the same way as for H-like ions in Table~\ref{H-like}.

In Table~\ref{Li-total} we present the total theoretical values for the $g$ factor of Li-like lead and uranium ions.
Except for the recoil corrections, all other contributions have been taken from 
the previous compilations \cite{vol14a,gla04,gla06,gla10}. 
Compared to Ref. \cite{vol14a}, we have strongly reduced the second largest theoretical
uncertainty, which was due to the nuclear recoil effect.
As one can see from Table~\ref{Li-total}, the QED  recoil effect is masked by uncertainties caused by
the nuclear size and polarization contributions.
The uncertainty of the nuclear size contribution is estimated as a quadratic sum
of the uncertainty which is due to the rms radius error bar \cite{ang13}
and the difference between the results obtained for the Fermi  and sphere nuclear 
charge distribution models. This is a rather conservative estimate of the 
uncertainty. It can be substantially reduced, provided 
the nuclear charge distribution parameters are known to a good accuracy
from experiments with the corresponding muonic atoms.
The more fundamental accuracy limit is actually set by the nuclear polarization uncertainty.
To reduce the uncertainty due to the nuclear effects, in Ref. \cite{sha:2002:062104}
it was proposed to study a specific difference between the $g$ factors of  Li- and H-like ions, 
\begin{eqnarray}
g^{\prime}= g_{(1s)^2 2s}- \xi g_{1s},
\end{eqnarray}   
where the parameter $\xi$ must be chosen to cancel the nuclear size correction
in this difference. It can be shown that both the  parameter $\xi$ and this
difference are very stable with respect to variations of the nuclear parameters
and nuclear models \cite{sha:2002:062104,sha06}. 

In case of lead  one obtains $\xi=0.1670264$ \cite{sha:2002:062104}. 
The  replacement of the Fermi model of the nuclear charge distribution   
with the sphere model changes the specific difference  $g^{\prime}$ 
by about $1\times 10^{-9}$.
But, as is mentioned above, this is a very conservative
estimate of the uncertainty. If, instead, we consider a variation
of the root-mean-square charge radius of the nucleus within
its double error bar, we get the change of  $g^{\prime}$ by about
$0.1\times 10^{-9}$ only.
The nuclear polarization correction
contributes  $-0.13(6)\times 10^{-9}$ to this difference  \cite{nef02,vol14}. 
At the same time, the QED recoil
contribution to the specific difference amounts to  $8.7\times 10^{-9}$.
This means that tests of the QED recoil effect on the $g$ factor of heavy ions
are possible on a few-percent level, provided all QED and electron-electron interaction  
corrections are calculated to the required accuracy.

\section{Conclusion}
In this paper we have evaluated the nuclear recoil effect on the
$g$ factor of H- and Li-like lead and uranium  ions for the finite-size-nucleus potential
and for the effective potentials which partially account for the electron-electron interaction
effects in Li-like ions. As the result, the second largest uncertainty
in the theoretical values of the $g$ factor of  $^{208}$Pb$^{79+}$ and $^{238}$U$^{89+}$
is strongly reduced.
The contribution of the QED recoil effect to the specific difference
of the $g$ factors of H- and Li-like ions is compared with the 
uncertainties due to the nuclear size and polarization effects. It is
shown that the QED recoil effect on the $g$ factor can be probed in 
experiments with heavy ions. This provides a unique opportunity
for tests of QED in a new region --- strong-coupling regime  beyond the Furry
picture.


\section{Acknowledgments}

This work was supported by the Russian Science Foundation (Grant No. 17-12-01097).
%
%

\begin{table}
\caption{The nuclear recoil contribution to the $1s$ $g$ factor
of H-like lead and uranium ions,
expressed in terms of the function $F(\alpha Z)$ defined by
Eq. (\ref{F}).
The  uncertainties are mainly
due to the approximate treatment of the nuclear size correction to the recoil effect (see the text).
}
\label{H-like}
\begin{tabular}{l@{\qquad}l@{\quad}l} \hline
Contribution &$^{208}$Pb$^{81+}$  & $^{238}$U$^{91+}$      \\
\hline
   Non-QED recoil, point nucleus  & 0.9632  &  0.9504    \\
   Non-QED recoil, extended nucleus  & 0.8746   &  0.7583    \\
   QED recoil, point nucleus  &0.8619    &1.4456      \\
   QED recoil, extended nucleus  &0.8564    &1.3491  \\
\hline
   Total recoil, point nucleus  &1.8251    & 2.3961     \\
 Total recoil, extended nucleus  &1.731(15)    & 2.107(61)   \\
\hline

\end{tabular}
\end{table}

\begin{table}
\caption{The nuclear recoil contribution to the $2s$ $g$ factor
of Li-like lead and uranium ions,
expressed in terms of the function $F(\alpha Z)$ defined by
Eq. (\ref{F}). All values are calculated for the extended nucleus case.
The total uncertainties account for the approximate treatment of the 
electron-electron interaction and
the nuclear size  effects (see the text).
 }
\label{Li-like}
\begin{tabular}{l@{\qquad}l@{\quad}l} \hline
Contribution &$\;\;$ $^{208}$Pb$^{79+}$  &$\;\;$ $^{238}$U$^{89+}$ \\
\hline
   Non-QED one-electron recoil &  $\;\,\,\,0.2597$   & $\;\,\,\,0.2471$      \\
   Non-QED two-electron recoil &  $-0.0072$   & $-0.0061$       \\   
\hline
   Total  non-QED recoil &  $\;\,\,\,0.2525$    &     $\;\,\,\, 0.2410$     \\
\hline
   QED recoil, Coulomb potential &  $\;\,\,\,0.1585$    &  $\;\,\,\,0.2693$    \\
   QED recoil, CH potential  &  $\;\,\,\,0.1525$   &  $\;\,\,\,0.2598$         \\
   QED recoil, LDF potential  & $\;\,\,\,0.1523$    &  $\;\,\,\,0.2597$       \\
   QED recoil, PZ potential  &  $\;\,\,\,0.1539$   & $\;\,\,\,0.2622$       \\
\hline
   Total recoil  &  $\;\,\,\,0.405(5)$    &   $\;\,\,\,0.501(17)$     \\
\hline

\end{tabular}
\end{table}

\begin{table}
\caption{The individual contributions to the ground-state $g$ factor of heavy Li-like ions.}
\label{Li-total}
\begin{center}
\begin{tabular}{l@{\qquad}l@{\quad}l}
                         &$\;\;\;\;\;\;^{208}$Pb$^{79+}$ &$\;\;\;\;\;\;^{238}$U$^{89+}$ \\ \hline
  Dirac value (point nucleus)   &   $\;\,\,\,1.932 002 904$ &  $\;\,\,\,1.910 722 624$   \\
  Finite nuclear size     &    $\;\,\,\,0.000 078 57(14)$ &  $\;\,\,\,0.000 241 62(36)$ \\
  One-electron QED  &  $\;\,\,\,0.002 408 1(5)$ &  $\;\,\,\,0.002 442 7(8)$ \\
  Screened QED   &  $-0.000 001 91(4)$ &  $-0.000 002 18(6)$ \\
 Interelectronic interaction  &   $\;\,\,\,0.002 139 34(4)$ &  $\;\,\,\,0.002 500 05(6)$  \\
  Non-QED  recoil   &  $\;\,\,\,0.000 000 239(2)$   &   $\;\,\,\,0.000 000 250(8)$   \\
  QED  recoil   &  $\;\,\,\,0.000 000 144(3)$   &  $\;\,\,\,0.000 000 270(10)$  \\
  Nuclear polarization       &$-0.000 000 04(2)$   & $-0.000 000 27(14)$   \\
  Total theory &  $\;\,\,\,1.936 627 3(5)$   &   $\;\,\,\,1.915 905 1(9)$\\
  Total theory from Ref.  \cite{vol14a}            & $\;\,\,\, 1.936 627 2(6)$ &  $\;\,\,\,1.915 904 8(11)$\\
 \hline
\end{tabular}
\end{center}
\end{table}




\begin{thebibliography}{99}


\bibitem[T. St\"ohlker (2000)]{sto00} 
Th.{~}St\"ohlker, P.H.{~}Mokler, F.{~}Bosch, R.W.{~}Dunford,
  F.{~}Franzke, O.{~}Klepper, C.{~}Kozhuharov, T.{~}Ludziejewski, F.{~}Nolden,
  H.{~}Reich, P.{~}Rymuza, Z.{~}Stachura, M.{~}Steck, P.{~}Swiat, and A.{~}Warczak,
\newblock Phys. Rev. Lett. {\bf 85},~3109 (2000).


\bibitem[A. Gumberidze (2005)]{gum05} 
A.{~}Gumberidze, Th.{~}St\"ohlker, D.{~}Bana\'{s}, K.{~}Beckert, P.{~}Beller, H.F.{~}Beyer, F.{~}Bosch,
  S.{~}Hagmann, C.{~}Kozhuharov, D.{~}Liesen, F.{~}Nolden, X.{~}Ma,
  P.H.{~}Mokler, M.{~}Steck, D.{~}Sierpowski, and S.{~}Tashenov,
\newblock Phys. Rev. Lett. {\bf 94},~223001 (2005).


\bibitem{sch91}
J.{~}Schweppe, A.{~}Belkacem, L.{~}Blumenfeld, N.{~}Claytor, B.{~}Feinberg,
  H.{~}Gould, V.E.{~}Kostroun, L.{~}Levy, S.{~}Misawa, J.R.{~}Mowat, and M.H.{~}Prior,
\newblock Phys. Rev. Lett. {\bf 66},~1434 (1991).


\bibitem{bra03}
C.{~}Brandau, C.{~}Kozhuharov, A.{~}M\"uller, W.{~}Shi, S.{~}Schippers,
  T.{~}Bartsch, S.{~}B\"ohm, C.{~}B\"ohme, A.{~}Hoffknecht, H.{~}Knopp,
  N.{~}Gr\"un, W.{~}Scheid, T.{~}Steih, F.{~}Bosch, B.{~}Franzke,
  P.H.{~}Mokler, F.{~}Nolden, M.{~}Steck, Th.{~}St\"ohlker, and
  Z.{~}Stachura,
\newblock Phys. Rev. Lett. {\bf 91},~073202 (2003).


\bibitem[P. Beiersdorfer (2005)]{bei05} 
P.{~}Beiersdorfer, H.{~}Chen, D.B.{~}Thorn, and E.{~}Tr\"abert,
\newblock Phys. Rev. Lett. {\bf 95},~233003 (2005).


\bibitem[D.A. Glazov (2011)]{gla11} 
D.A.{~}Glazov, Y.S.{~}Kozhedub, A.V.{~}Maiorova, V.M.{~}Shabaev,
  I.I.{~}Tupitsyn, A.V.{~}Volotka, C.{~}Kozhuharov, G.{~}Plunien, and Th.{~}St\"ohlker,
\newblock Hyp. Interact. {\bf 199},~71 (2011).


\bibitem[A.V. Volotka (2013)]{vol13} 
A.V.{~}Volotka, D.A.{~}Glazov, G.{~}Plunien, and V.M.{~}Shabaev,
\newblock Ann. Phys. (Berlin) {\bf 525},~636 (2013).


\bibitem{kla94} 
I.{~}Klaft, S.{~}Borneis, T.{~}Engel, B.{~}Fricke, R.{~}Grieser, G.{~}Huber,
  T.{~}K\"uhl, D.{~}Marx, R.{~}Neumann, S.{~}Schr\"oder, P.{~}Seelig, and
  L.{~}V\"olker,
\newblock Phys. Rev. Lett. {\bf 73},~2425 (1994).


\bibitem{cre96} 
J.R.{~}Crespo L\'opez-Urrutia, P.{~}Beiersdorfer, D.W.{~}Savin, and K.{~}Widmann,
\newblock Phys. Rev. Lett. {\bf 77},~826 (1996).


\bibitem{cre98}
J.R.{~}Crespo L\'opez-Urrutia, P.{~}Beiersdorfer, K.{~}Widmann, B.B.{~}Birkett,
  A.-M.{~}M\aa{}rtensson-Pendrill, and M.G.H.{~}Gustavsson,
\newblock Phys. Rev. A {\bf 57},~879 (1998).


\bibitem{see98} 
P.{~}Seelig, S.{~}Borneis, A.{~}Dax, T.{~}Engel, S.{~}Faber, M.{~}Gerlach,
  C.{~}Holbrow, G.{~}Huber, T.{~}K\"uhl, D.{~}Marx, K.{~}Meier, P.{~}Merz,
  W.{~}Quint, F.{~}Schmitt, M.{~}Tomaselli, L.{~}V\"olker, H.{~}Winter,
  M.{~}W\"urtz, K.{~}Beckert, B.{~}Franzke, F.{~}Nolden, H.{~}Reich,
  M.{~}Steck, and T.{~}Winkler,
\newblock Phys. Rev. Lett. {\bf 81},~4824 (1998).


\bibitem{bei01}
P.{~}Beiersdorfer, S.B.{~}Utter, K.L.{~}Wong, J.R.{~}Crespo L\'opez-Urrutia,
  J.A.{~}Britten, H.{~}Chen, C.L.{~}Harris, R.S.{~}Thoe, D.B.{~}Thorn,
  E.{~}Tr\"abert, M.G.H.{~}Gustavsson, C.{~}Forss\'en, and A.-M.{~}M\aa{}rtensson-Pendrill,
\newblock Phys. Rev. A {\bf 64},~032506 (2001).


\bibitem{ull15} 
J.{~}Ullmann, Z.{~}Andelkovic, A.{~}Dax, W.{~}Geithner, C.{~}Geppert,
  C.{~}Gorges, M.{~}Hammen, V.{~}Hannen, S.{~}Kaufmann, K.{~}K{\"o}nig, Yu.{~}Litvinov, 
  M.{~}Lochmann, B.{~}Maass, J.{~}Meisner, T.{~}Murb{\"o}ck,
  R.{~}S\'anchez, M.{~}Schmidt, S.{~}Schmidt,
  M.{~}Steck, Th.{~}Stöhlker, R.C.{~}Thompson, J.{~}Vollbrecht,
  C.{~}Weinheimer, and W.{~}N{\"o}rtersh{\"a}user,
\newblock J. Phys. B: At. Mol. Opt. Phys. {\bf 48},~144022 (2015).


\bibitem[V.M. Shabaev (2001)]{sha01} 
V.M.{~}Shabaev, A.N.{~}Artemyev, V.A.{~}Yerokhin, O.M.{~}Zherebtsov, and G.{~}Soff,
\newblock Phys. Rev. Lett. {\bf 86},~3959 (2001).


\bibitem[J. Ullmann (2017)]{ull17}  
J.{~}Ullmann, Z.{~}Andelkovic, C.{~}Brandau, A.{~}Dax, C.{~}Geithner, W.
  amd~Geppert, C.{~}Gorges, M.{~}Hammen, V.{~}Hannen, S.{~}Kaufmann,
  K.{~}K{\"o}nig, Yu.A.{~}Litvinov, M.{~}Lochmann, B.{~}Maa{\ss},
  J.{~}Meisner, T.{~}Murb{\"o}ck, R.{~}S\'anchez, M.{~}Schmidt, 
  S.{~}Schmidt, M.{~}Steck, Th.{~}St{\"o}hlker, R.C.{~}Thompson, C.{~}Trageser, 
  J.{~}Vollbrecht, C.{~}Weinheimer, and W.{~}N{\"o}rtersh{\"a}user,
\newblock Nat. Commun. {\bf 8},~15484 (2017).


\bibitem{volotka:12:prl}
A.V.{~}Volotka, D.A.{~}Glazov, O.V.{~}Andreev, V.M.{~}Shabaev, I.I.{~}Tupitsyn, and G.{~}Plunien,
\newblock Phys. Rev. Lett. {\bf 108},~073001 (2012).


\bibitem{kar17}
J.-P.{~}Karr,
\newblock Nature Physics {\bf 13},~533 (2017).


\bibitem{haf00}
H.{~}H\"affner, T.{~}Beier, N.{~}Hermanspahn, H.J.{~}Kluge, W.{~}Quint,
  S.{~}Stahl, J.{~}Verd\'u, and G.{~}Werth,
\newblock Phys. Rev. Lett. {\bf 85},~5308 (2000).


\bibitem{ver04}
J.{~}Verd\'u, S.{~}Djeki\'c, S.{~}Stahl,
  T.{~}Valenzuela, M.{~}Vogel, G.{~}Werth, T.{~}Beier, H.J.{~}Kluge, and W.{~}Quint,
\newblock Phys. Rev. Lett. {\bf 92},~093002 (2004).


\bibitem{stu11}
S.{~}Sturm, A.{~}Wagner, B.{~}Schabinger, J.{~}Zatorski, Z.{~}Harman,
  W.{~}Quint, G.{~}Werth, C.H.{~}Keitel, and K.{~}Blaum,
\newblock Phys. Rev. Lett. {\bf 107},~023002 (2011).


\bibitem{wag13}
A.{~}Wagner, S.{~}Sturm, F.{~}K\"ohler, D.A.{~}Glazov, A.V.{~}Volotka,
  G.{~}Plunien, W.{~}Quint, G.{~}Werth, V.M.{~}Shabaev, and K.{~}Blaum,
\newblock Phys. Rev. Lett. {\bf 110},~033003 (2013).


\bibitem{lin13}
D.{~}von{~}Lindenfels, M.{~}Wiesel, D.A.{~}Glazov, A.V.{~}Volotka,
  M.M.{~}Sokolov, V.M.{~}Shabaev, G.{~}Plunien, W.{~}Quint, G.{~}Birkl,
  A.{~}Martin, and M.{~}Vogel,
\newblock Phys. Rev. A {\bf 87},~023412 (2013).


\bibitem{stu13}
S.{~}Sturm, A.{~}Wagner, M.{~}Kretzschmar, W.{~}Quint, G.{~}Werth, and K.{~}Blaum,
\newblock Phys. Rev. A {\bf 87},~030501(R) (2013).


\bibitem{stu14}
S.{~}Sturm, F.{~}K\"ohler, J.{~}Zatorski, A.{~}Wagner, Z.{~}Harman, G.{~}Werth,
  W.{~}Quint, C.H.{~}Keitel, and K.{~}Blaum,
\newblock Nature {\bf 506},~467 (2014).


\bibitem{koel16}
F.{~}K\"ohler, K.{~}Blaum, M.{~}Block, S.{~}Chenmarev, S.{~}Eliseev,
  D.A.{~}Glazov, M.{~}Goncharov, J.{~}Hou, A.{~}Kracke, D.A.{~}Nesterenko,
  Yu.N.{~}Novikov, W.{~}Quint, E.M.{~}Ramirez, V.M.{~}Shabaev,
  S.{~}Sturm, A.V.{~}Volotka, and G.{~}Werth,
\newblock Nat Commun. {\bf 7},~10246 (2016).


\bibitem{shaxx}
V.M. Shabaev, D.A. Glazov, A.V. Malyshev, and I.I. Tupitsyn, to be published.


\bibitem{sha02b}
V.M.{~}Shabaev and V.A.{~}Yerokhin,
\newblock Phys. Rev. Lett. {\bf 88},~091801 (2002).


\bibitem{nef02}
A.V.{~}Nefiodov, G.{~}Plunien, and G.{~}Soff,
\newblock Phys. Rev. Lett. {\bf 89},~081802 (2002).


\bibitem{sha:2002:062104}
V.M.{~}Shabaev, D.A.{~}Glazov, M.B.{~}Shabaeva, V.A.{~}Yerokhin, G.{~}Plunien, and G.{~}Soff,
\newblock Phys. Rev. A {\bf 65},~062104 (2002).


\bibitem[V.M. Shabaev (2006)]{sha06}
V.M.{~}Shabaev, D.A.{~}Glazov, N.S.{~}Oreshkina, A.V.{~}Volotka, G.{~}Plunien,
  H.J.{~}Kluge, and W.{~}Quint,
\newblock Phys. Rev. Lett. {\bf 96},~253002 (2006).


\bibitem{vol14}
A.V.{~}Volotka and G.{~}Plunien,
\newblock Phys. Rev. Lett. {\bf 113},~023002 (2014).


\bibitem{yerokhin:2016:100801}
V.A.{~}Yerokhin, E.{~}Berseneva, Z.{~}Harman, I.I.{~}Tupitsyn, and C.H.{~}Keitel,
\newblock Phys. Rev. Lett. {\bf 116},~100801 (2016).


\bibitem{sha01a}
V.M.{~}Shabaev,
\newblock Phys. Rev. A {\bf 64},~052104 (2001).


\bibitem{sha02a}
V.M.{~}Shabaev,
\newblock Phys. Rep. {\bf 356},~119 (2002).


\bibitem{shc15}
A.A.{~}Shchepetnov, D.A.{~}Glazov, A.V.{~}Volotka, V.M.{~}Shabaev,
  I.I.{~}Tupitsyn, and G.{~}Plunien,
\newblock J. Phys. Conf. Ser. {\bf 583},~012001 (2015).


\bibitem{sha04}
V.M.{~}Shabaev, I.I.{~}Tupitsyn, V.A.{~}Yerokhin, G.{~}Plunien, and G.{~}Soff,
\newblock Phys. Rev. Lett. {\bf 93},~130405 (2004).


\bibitem{sap96}
J.{~}Sapirstein and W.R.{~}Johnson,
\newblock J. Phys. B: At. Mol. Opt. Phys. {\bf 29},~5213 (1996).


\bibitem{ang13}
I.{~}Angeli and K.P.{~}Marinova,
\newblock At. Data Nucl. Data Tables {\bf 99},~69 (2013).


\bibitem{per81}
J.P.{~}Perdew and A.{~}Zunger,
\newblock Phys. Rev. B {\bf 23},~5048 (1981).


\bibitem{sap02}
J.{~}Sapirstein and K.T.{~}Cheng,
\newblock Phys. Rev. A {\bf 66},~042501 (2002).


\bibitem{sha05}
V.M.{~}Shabaev, I.I.{~}Tupitsyn, K.{~}Pachucki, G.{~}Plunien, and V.A.{~}Yerokhin,
\newblock Phys. Rev. A {\bf 72},~062105 (2005).


\bibitem{sap11}
J.{~}Sapirstein and K.T.{~}Cheng,
\newblock Phys. Rev. A {\bf 83},~012504 (2011).


\bibitem{mal17}
A.V.{~}Malyshev, D.A.{~}Glazov, A.V.{~}Volotka, I.I.{~}Tupitsyn,
  V.M.{~}Shabaev, G.{~}Plunien, and Th.{~}St\"ohlker,
\newblock Phys. Rev. A {\bf 96},~022512 (2017).


\bibitem{ale15}
I.A.{~}Aleksandrov, A.A.{~}Shchepetnov, D.A.{~}Glazov, and V.M.{~}Shabaev,
\newblock J. Phys. B: At. Mol. Opt. Phys. {\bf 48},~144004 (2015).


\bibitem{vol14a}
A.V.{~}Volotka, D.A.{~}Glazov, V.M.{~}Shabaev, I.I.{~}Tupitsyn, and G.{~}Plunien,
\newblock Phys. Rev. Lett. {\bf 112},~253004 (2014).


\bibitem{gla04}
D.A.{~}Glazov, V.M.{~}Shabaev, I.I.{~}Tupitsyn, A.V.{~}Volotka,
  V.A.{~}Yerokhin, G.{~}Plunien, and G.{~}Soff,
\newblock Phys. Rev. A {\bf 70},~062104 (2004).


\bibitem{gla06}
D.A.{~}Glazov, A.V.{~}Volotka, V.M.{~}Shabaev, I.I.{~}Tupitsyn, and G.{~}Plunien,
\newblock Phys. Lett. A {\bf 357},~330 (2006).


\bibitem{gla10}
D.A.{~}Glazov, A.V.{~}Volotka, V.M.{~}Shabaev, I.I.{~}Tupitsyn, and G.{~}Plunien,
\newblock Phys. Rev. A {\bf 81},~062112 (2010).


\end{thebibliography}
\end{document}